\documentstyle[rawfonts,budapest]{article}  
\frompage{000} \topage{000}                                              
\def\rightmark{The decoupling problem at RHIC} 
\def\leftmark{D. Moln\'ar et al.}
 
\newcommand\be{\begin{equation}}
\newcommand\bea{\begin{eqnarray}}
\newcommand\ee{\end{equation}}
\newcommand\eea{\end{eqnarray}}

\newcommand\vx{\vec x}
\newcommand\vq{\vec q}
\newcommand\vK{\vec K}
\newcommand\vp{\vec p}
\newcommand\SKIP[1]{}

\title{The decoupling problem at RHIC}
\authors{
{D\'enes Moln\'ar$^1$ and Miklos Gyulassy$^{1}$ %
}\\[2.812mm]
{\normalsize
\hspace*{-8pt}$^1$ Physics Department, Columbia University\\
         538 W. 120th Street, New York, NY 10027 \\[0.2ex] 
}}
 
\abstract{
We investigate whether it is possible 
to {\em dynamically} generate from classical transport theory
the observed surprising $R_{out} \approx R_{side}$ 
in Au+Au at $\sqrt{s} = 130A$ GeV at RHIC \cite{starHBT,phenixHBT}.
We obtained covariant solutions to the Boltzmann transport equation 
via the MPC technique \cite{v2},
for a wide range of partonic initial conditions and opacities.
We demonstrate that there exist 
transport solutions that yield a freezeout distribution
with $R_{out} < R_{side}$ for $K_\perp > \sim 1.5$ GeV.
These solutions correspond to continuous evaporation-like freezeout, 
where the emission duration is {\em comparable} to the source size.
Naively this would mean $R_{out} > R_{side}$.
Nevertheless, our sources exhibit  $R_{out} < R_{side}$
because they are narrower in the 'out' than in the 'side' direction
and, in addition, a positive $x_{out}-t$ correlation develops reducing $R_{out}$ 
further.
}

 
\begin{document}
 
\maketitle
\setcounter{page}{1}

\section{Introduction}\label{intro}
Long-standing theoretical interests of the RMKI KFKI research group
have been hydrodynamical, decoupling and transport phenomena%
\cite{ALCOR,Magas:1999yb,Magas:2000jx,TamasHBT}.
A recent theoretical puzzle is why
the pion source extracted via HBT interferometry
in Au+Au at $\sqrt{s} = 130A$ GeV at RHIC
has approximately identical 'out' and 'side' radii
for average pair momenta $K_\perp < 0.7$ GeV\cite{starHBT,phenixHBT}.
Most naively, this indicates very short pion emission duration,
corresponding to almost instantaneous decoupling.

The observed ratio $R_{out} / R_{side}\approx 0.9-1.1$
contradicts theoretical expectations
based on cylindrically symmetric and longitudinally boost invariant
ideal hydrodynamics,
indicating $R_{out} / R_{side}$ significantly
larger than one
\cite{DirkHBT,Uliradii}.
On the other hand,
a hydrodynamical approach to the decoupling problem is inherently questionable
because in the decoupling region the mean free path
exceeds the side of the system,
while hydrodynamics is only valid if the mean free path is zero.
As a workaround, hydrodynamics is usually supplemented
with an {\em arbitrary} freezeout prescription
postulating sudden freezeout on a three-dimensional
hypersurface,
which is 
typically parametrized by a constant freezeout temperature or energy density.
The unsatisfactory nature of this procedure has been realized recently
and several improvements have been proposed \cite{Magas:1999yb}.

A dynamical framework with a self-consistent freezeout is
provided by classical transport theory\cite{Yang,ZPC,nonequil,v2,finrange},
where the rate of interactions is given by the product
of the density and the cross section $\lambda^{-1} = n\sigma$.
As the system expands, it rarifies, and the rate of particle interactions
drops accordingly.
Because the theory is not limited to equilibrium dynamics, it 
also allows for systematic studies of dissipative effects.
Dissipation is crucial in order to reproduce\cite{v2} the observed
saturation of elliptic flow at $p_\perp > 2$ GeV\cite{starv2}.

The equilibrium assumption was partly relaxed in an earlier
study\cite{hydro+urqmd},
where hydrodynamical evolution was only assumed until hadronization
at which point particles were fed into a hadronic cascade.
The calculation gave $R_{out} / R_{side} \approx 1.2-1.4$,
still above one.
On the other hand,
it is not clear whether the hadronic densities were large enough
to support equilibrium on the freezeout hypersurface,
which would be necessary to justify the assumed equilibrium evolution until the
end of hadronization.
In fact, antiprotons are known to scatter only once on average
in this approach,
indicating the need for a nonequilibrium approach before hadronization.

In this study we investigate how nonequilibrium evolution in the partonic
phase influences the pion distributions at freezeout.
Utilizing the MPC numerical technique\cite{MPCsource},
we compute the covariant 
freezeout distributions
for a variety of initial conditions for Au+Au at $\sqrt{s} = 130 A$ GeV 
based on perturbative QCD, gluon saturation models, or a combination of both.
Varying the transport opacity,
which is an effective measure of the degree of collectivity\cite{v2},
we study under what conditions $R_{out} < R_{side}$ is possible.

\section{Two-particle HBT interferometry}\label{HBT}

Hanbury-Brown and Twiss interferometry 
is a truly remarkable technique
to deduce
{\em space-time} geometrical 
information purely via measuring momentum correlations.
We review here only a few essential features
and refer the reader to Refs. \cite{MiklosHBT,TamasHBT,UliHBT} and references 
therein.

Detailed studies of pion production by an ensemble of
classical currents\cite{MiklosHBT}
showed that for a chaotic boson source $\rho(\vx, t)$,
in the limit of a large particle numbers,
and also neglecting final state interactions,
the two-particle correlation function is 
given by the {\em space-time} Fourier transform of the source
as 
\be
C(q,K) \equiv
C(p_1, p_2) \equiv \frac{N(p_1, p_2)}{N(p_1) N(p_2)} =
1 +  \left|\int d^4 x \, \rho(\vx, t) e^{i q x}\right|^2  \ .
\label{Eq:corrfn}
\ee
Here $q^\mu \equiv p_1^\mu - p_2^\mu$ is momentum difference, while 
$K^\mu \equiv (p_1^\mu + p_2^\mu) / 2$ is the average momentum of the
boson pair.
\SKIP{
Simply speaking,
the correlation function measures the size of
 the four-dimensional space-time
region where particles with momenta around $K^\mu$ were emitted
(regions of ``homogeneity'' by Sinyukov).
}

Unfortunately, Eq. (\ref{Eq:corrfn}) cannot be uniquely inverted
to obtain the source distribution,
because the phase information is lost
and, in addition,
only a three-dimensional 
projection of the four-dimensional correlation function is measurable
because $q^\mu$ is constrained\cite{UliHBT} by
\be
q^\mu K_\mu = (m_1^2 - m_2^2) / 2 = 0   
\qquad({\rm for\ identical\ particles}) \ .
\label{massshell}
\ee
Despite the inversion problem,
HBT measurements provide strong constraints 
on the freezeout distributions of particles and therefore on
the possible dynamical scenarios in heavy-ion collisions.

Conventionally,
$C(\vq,K)$ is expressed in terms of
the 'out-side-long' variables, $q_O$, $q_S$, and $q_L$. 
These are defined in the longitudinally boosted LCMS reference frame,
where $K_z = 0$ with $z$ or 'long' being the beam direction,
'out' the direction of $\vK_\perp$, and 'side' the one orthogonal to both.
In this frame, $K^\mu_{LCMS} = (\tilde K^0, K_\perp, 0, 0)$,
$\vq_{LCMS} = (q_O, q_S, q_L)$, 
$x^\mu_{LCMS} \equiv (\tilde t, x_O, x_S, x_L)$
and Eq. (\ref{massshell}) gives
\be
q^\mu x_\mu = -q_O (x_O - \tilde t K_\perp/\tilde K^0) - q_S x_S - q_L x_L \ .
\ee
Clearly,
the measurable projected correlation function is completely blind to
the dependence of the source function on the fourth orthogonal coordinate
$\tilde t + x_O K_\perp/\tilde K^0$.

Experimentally,
the measured correlation function
(after correcting for final state interactions)
is fitted with a Gaussian, which for central collisions ($b=0$) and 
at midrapidity is constrained by symmetry to be of the form
\be
C(\vq, K) = \lambda(K) \exp\!\left[ -q_O^2 R_O^2(K) - q_S^2 R_S^2(K) - q_L^2 R_L^2(K)\right] \ .
\ee
Here $R_O$, $R_S$ and $R_L$ are the 'out', 'side', and 'long' HBT radii.
For a {\em perfectly Gaussian} source,
the correlation function is Gaussian and 
\bea
R_O^2(K) &=& \langle \Delta x_O^2 \rangle_K
        + v_\perp^2 \langle \Delta \tilde t^2 \rangle_K
        - 2 v_\perp  \langle \Delta x_O \Delta \tilde t \rangle_K
\nonumber \\
R_S^2(K) &=& \langle \Delta x_S^2 \rangle_K, \qquad{\rm and} \quad R_L^2(K) = \langle \Delta x_L^2\rangle_K \ ,
\label{HBTradii}
\eea
where $v_\perp \equiv K_\perp / \tilde K^0$.
Thus $R_S$ and $R_L$ have simple geometric interpretation 
as the 'side' and 'long' widths of the distribution function,
while $R_O$ is a mixture of the 'out' width, time spread, {\em and}
the $x_O-\tilde t$ correlation.

\section{Covariant parton transport theory}\label{transport}

\subsection{Transport equation}
We consider here, as in Refs. 
\cite{Yang,ZPC,nonequil,v2},
the simplest but nonlinear
form of Lorentz-covariant Boltzmann transport theory
in which the on-shell phase space density $f(x,\vp)$,
evolves with an elastic $2\to 2$ rate as
\be
p_1^\mu \partial_\mu f_1 = S(x, \vp_1) + \int\limits_2\!\!\!\!
\int\limits_3\!\!\!\!
\int\limits_4\!\!
\left(
f_3 f_4 - f_1 f_2
\right)
W_{12\to 34} \delta^4(p_1+p_2-p_3-p_4) \ .
\label{Eq:Boltzmann_22}
\ee
Here $W$ is the square of the scattering matrix element,
the integrals are shorthands
for $\int\limits_i \equiv \int \frac{g\ d^3 p_i}{(2\pi)^3 E_i}$,
where $g$ is the number of internal degrees of freedom,
while $f_j \equiv f(x, \vp_j)$.
The initial conditions are specified by the source function $S(x,\vp)$.
For our applications below,
we  neglect quark degrees of freedom and
interpret  $f(x,\vp)$ as describing
an ultrarelativistic massless gluon gas 
with $g=16$ (8 colors, 2 helicities).

\SKIP{
Eq. (\ref{Eq:Boltzmann_22}) can be easily extended
to include inelastic matrix elements,
such as $gg \leftrightarrow ggg$,
and proper Bose and/or Fermi statistics can be introduced as well.
However, at present there is no practical algorithm
to compute accurate numerical solutions to the resulting transport equations.
Therefore, we limit our study to classical particles with elastic interactions.
}

The elastic gluon scattering matrix elements in dense parton systems
were modeled with the isotropic form $ d\sigma_{el}/dt = \sigma_0 (s) / s$.
The simplified angular dependence
is justified by our previous study\cite{v2},
where we showed that the covariant transport solutions
do not depend {\em explicitly}
on the differential cross section
but only on the {\em transport opacity}
\be
\chi\equiv \frac{\sigma_{tr}}{\sigma_{el}} \langle n \rangle
\approx \sigma_{tr}
\langle \int dz  
\rho\left({\bf x}_0+ z\hat{\bf n},\tau=\frac{z}{c} \right)\rangle
\;\; .
\label{Eq:tropacity}
\ee
Here
\be
\sigma_{tr}(s) \equiv \int d\sigma_{el} \sin^2\theta_{cm}=\int
dt\, \frac{d\sigma_{el}}{dt}\frac{4t}{s}\left(1-\frac{t}{s}\right)
\ee
is the transport cross section (in our case, $\sigma_{tr} = 2\sigma_0/3$), while $\langle n \rangle$
is the average number of scatterings a parton undergoes.
We neglected the weak logarithmic energy dependence
of the total cross section and took a constant $\sigma_0$ for 
simplicity.
For a fixed nuclear geometry,
a given transport opacity $\chi$ represents
a whole class of initial conditions and partonic matrix elements,
which is demonstrated by the approximate proportionality\cite{v2}
$\chi\propto \sigma_{tr} dN_g(\tau_0)/d\eta$.

We solved Eq. (\ref{Eq:Boltzmann_22}) numerically
via the MPC parton cascade algorithm \cite{MPCsource},
which utilizes 
the parton subdivision technique\cite{Yang,ZPC}
to obtain the correct, covariant solutions to the transport equation.
Parton subdivision is essential to eliminate numerical artifacts caused by
acausal (superluminal)
propagation due to
action at a distance\cite{nonequil}.
For initial partonic densities expected at RHIC,
the severe violation of Lorentz covariance
in the naive cascade algorithm that employs no subdivision
artificially reduces elliptic flow and heats up the
$p_\perp$ spectra\cite{v2,finrange}.

\SKIP{
\subsection{Hadronization schemes}

To compare to the data, the partons must be hadronized.
We applied two different hadronization prescriptions.
The first one was motivated by parton-hadron duality,
where as in Ref. \cite{EKRT}, we 
assumed that each gluon gets converted
to a pion with equal probability for the three isospin states.
This means that 
$f_{h^-}(\vec p_\perp) \approx f_{\pi^-}(\vec p_\perp)
= f_{g}(\vec p_\perp) / 3$.

The other scheme was independent fragmentation,
where we considered only the $g\to \pi^{\pm}$ channel
with the next-to-leading-order
fragmentation function taken from
 Ref. \cite{BKK95}.
We took
the scale factor $s \equiv \log(Q^2)/\log(Q_0^2)$ to be zero
because our initial gluon distribution taken from HIJING was already
``self-quenched''
due to initial and final state radiation.
}

\subsection{Initial conditions}

We modeled central Au+Au collisions at RHIC 
with three different classes of initial conditions,
corresponding to different assumptions on the initial transverse momentum
distribution and density distribution of partons.
In all three cases, the evolution started from a 
longitudinally boost invariant Bjorken tube 
at proper time $\tau_0=0.1$~fm/$c$,
with isotropic momentum distribution in the local rest frame
and 
with uniform  pseudorapidity $\eta\equiv 1/2 \log[(t+z)/(t-z)]$ distribution
between $|\eta| < 5$.

The first class was essentially the same as the minijet initial conditions 
in Ref. \cite{v2}.
The initial transverse density distribution was
proportional to the binary collision distribution for 
two Woods-Saxon distributions,
while the $p_\perp$ distribution was a fit to the minijet distribution
predicted by HIJING\cite{HIJING} (without shadowing and jet quenching).
To include the transport effect
of the soft partonic component present in HIJING {\em in addition} to the minijets,
in the present study we increased the parton density by a factor five,
to yield $dN_g(\tau_0)/d\eta = 1000$.

The second class was based on gluon saturation models.
Here we assumed a uniform Bjorken cylinder with a radius $R_0 = 6$ fm,
$dN_g/d\eta = 1000$ and a
constant $d^2N/p_\perp dp_\perp$ distribution that vanishes for 
$p_\perp > Q_{sat} = 1.2$ GeV.

The third class was a combination of the first two. 
We took the minijet part with the original HIJING normalization 
($dN/d\eta = 210$) and added four times as many saturated soft gluons.

\section{Results}

Using MPC, we computed the freezeout distribution numerically 
for central Au+Au at 130$A$ GeV
as a function of the average
pair momentum and transport opacity, for all three classes of
initial conditions.

The freezeout distribution $d^4N/d^4 x$ was defined
as the distribution of space-time coordinates for 
the point of last interaction of the test particles,
with the strong assumption that this point is not affected by 
hadronization.
We furthermore neglected resonance contributions to the pion yield.
The same two hadronization schemes, parton-hadron duality and independent
fragmentation, were applied as in Ref. \cite{v2}.

Of our primary interest were the integrated distributions 
$d^2N/r dr d\tilde t$, 
the equivalent of the usual
hydrodynamical freezeout curves in the $r-\tau$ plane, 
and $d^2N/dx_O dx_S$, which is the source projected onto the 'out-side' plane.
We worked 
in the approximation $p_{1,\perp} \approx p_{2,\perp} \approx K_\perp$,
which is valid up to corrections of $O(|\vq|/|\vK|)$. 
To get a simple estimate of the radii, we used Eq. (\ref{HBTradii})
even though our sources are non-Gaussian.
A more detailed analysis of the radii based on the momentum-space correlation
function is in progress.

\begin{figure}[hbpt]
\epsfysize=13cm
\vspace*{-0.5cm}
\epsfbox{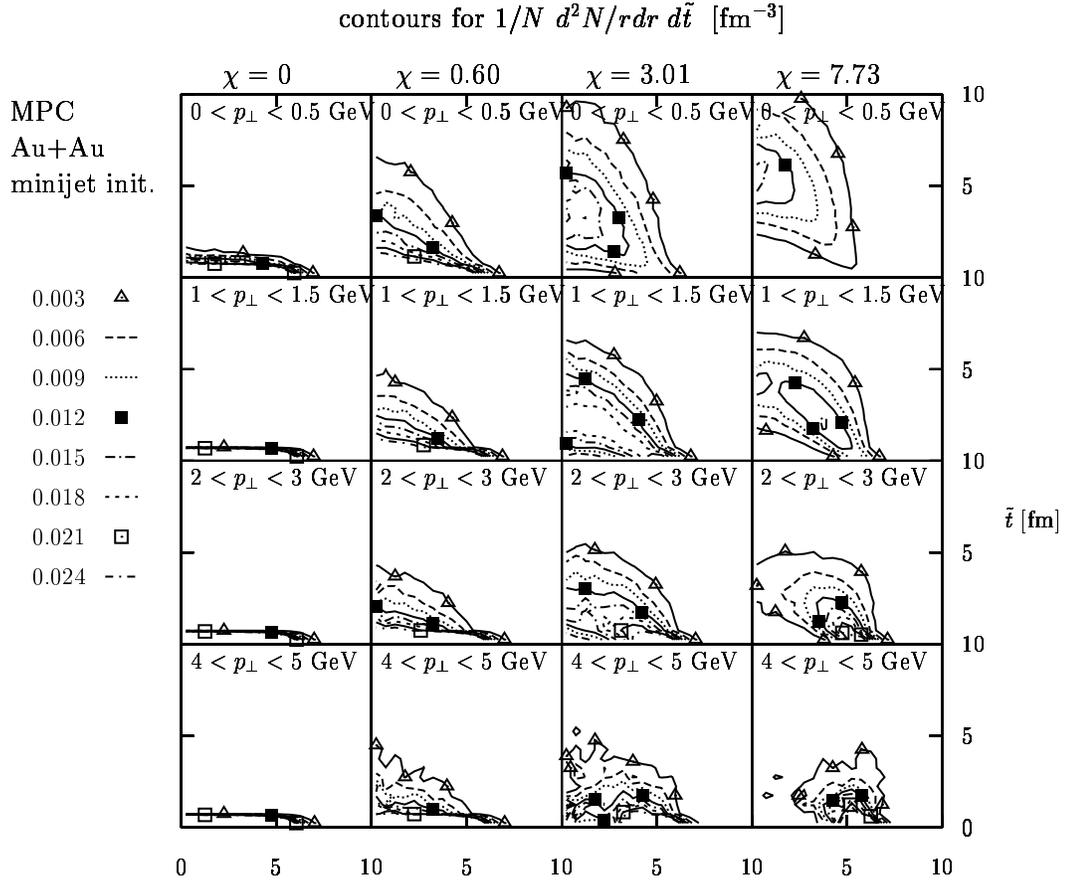}
\vspace*{-1 cm}
\caption[]{Freezeout $r-\tilde t$ distribution as a function of transport
opacity and $p_\perp$ for the minijet initial condition
with hadronization via parton-hadron duality.}
\label{fig:1}
\end{figure}

Figure \ref{fig:1} shows the freezeout distribution in $r-\tilde t$ 
coordinates ($r\equiv \sqrt{x_O^2 + x_S^2}$) as a function of the pair momentum
and transport opacity for the minijet initial condition with
hadronization via parton-hadron duality.
Unlike the sharp freezeout imposed in hydrodynamical models,
the transport theory freezeout is a continuous\cite{nonequil}, {\em evaporation-like} process.
For a given (nonzero) opacity,
the larger the $p_\perp$ of the particle,
the earlier it decouples and the closer it is to the surface of the nuclei.
Thus, low-$p_\perp$ partons freeze out in the center at late times,
while high-$p_\perp$ ones escape from the surface at early times.
Furthermore, 
a larger opacity increases decoupling times, especially for low-$p_\perp$ 
particles.
We observed the same features for the other two classes of initial conditions.

\begin{figure}[hbpt]
\epsfysize=13cm
\vspace*{-0.5cm}
\epsfbox{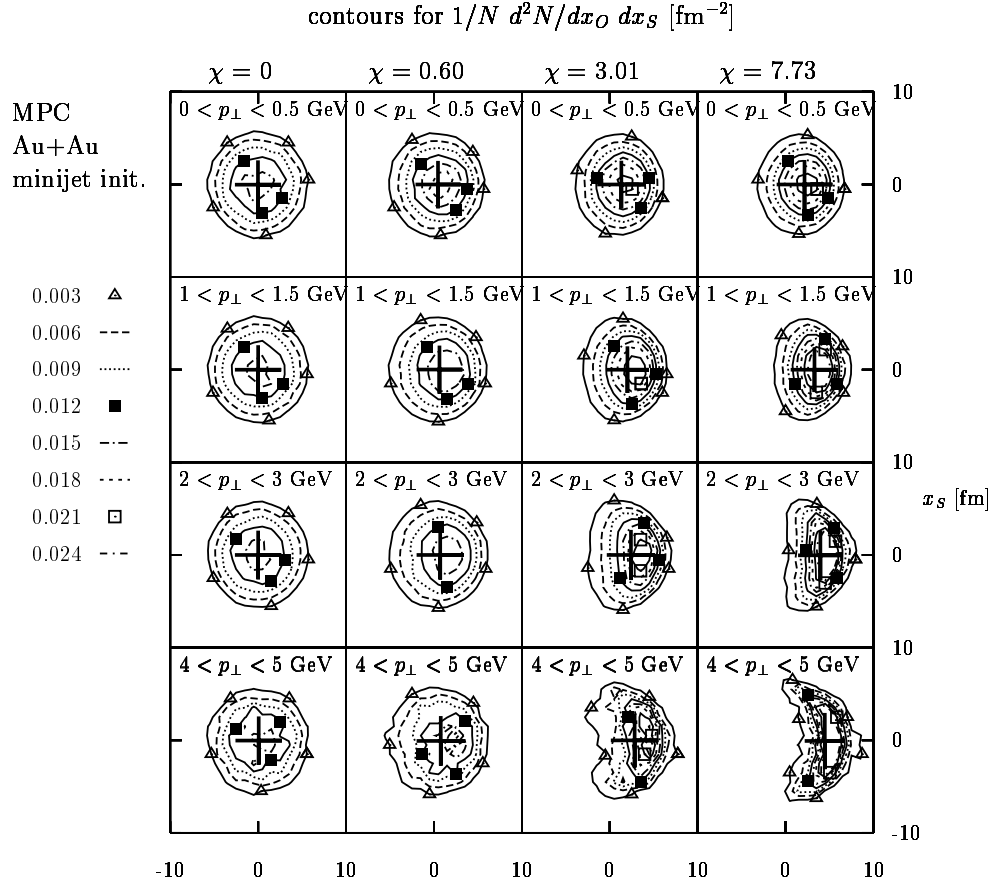}
\vspace*{-1cm}
\caption[]{Freezeout $x_O-x_S$ distribution as a function of transport
opacity and $K_\perp$ for the minijet initial condition 
with hadronization via parton-hadron duality.}
\label{fig:2}
\end{figure}

Figure \ref{fig:2} shows the shape of the source in $x_O-x_S$ 
coordinates as a function of the pair momentum
and transport opacity for the minijet initial condition
with hadronization via parton-hadron duality.
Results for the saturation and the 'combined' initial condition
were very similar. 
The black crosses are positioned at the first moment of the distributions
$(\langle x_O\rangle, \langle x_S \rangle)$,
while their extension corresponds to the square root of the second moments
$(\sqrt{\langle \Delta x_O^2 \rangle}, \sqrt{\langle \Delta x_S^2 \rangle})$.
As the opacity increases, the source shrinks in the 'out' direction
and the higher the $p_\perp$, the more pronounced the reduction of 
$\langle \Delta x_O^2\rangle$ is.
From Eq. (\ref{HBTradii}) it is clear that this will help reduce $R_O$,
at least if the other two terms do not change.
On the other hand,
we find that
the source size in the 'side' direction is approximately
independent of both the opacity and $p_\perp$.
Thus, surprisingly, 
$R_S$ is unaffected by the strong collective dynamics.

The mean $\langle x_O\rangle$ does not coincide 
with the maximum of the distributions showing that the distributions
are asymmetric in the 'out' direction,
i.e.,
they are clearly non-Gaussian.
In fact, the change in the shape of contour lines with increasing opacity
is much more pronounced than the slow decrease of the second moment 
$\langle \Delta x_O^2\rangle$
because there is a long tail in the negative 'out' direction.
We expect that $R_O$ extracted via a least-squared Gaussian fit
would turn out to be somewhat {\em smaller} than from Eq. (\ref{HBTradii}).

\begin{figure}[h]
\epsfysize=12cm
\epsfbox{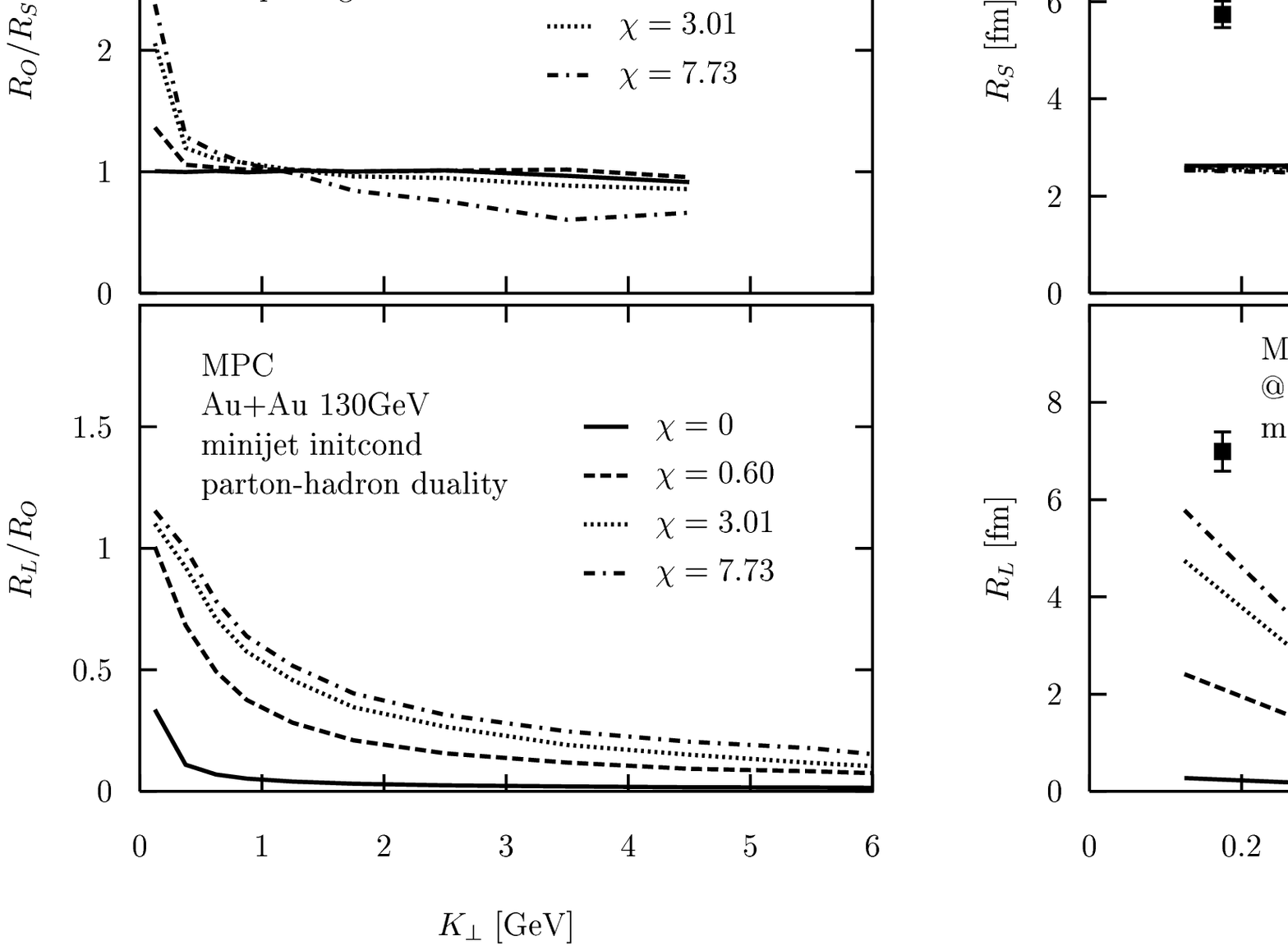}
\vspace*{-1cm}
\caption[]{HBT radii (right column) and ratios of radii (left column) 
as a function of transport
opacity and $K_\perp$ for the minijet initial condition.
with hadronization via parton-hadron duality (top and middle left)
or independent fragmentation (bottom left and whole right column).}
\label{fig:3}
\end{figure}

The left column in Fig. \ref{fig:3} 
shows the dependence of the $R_O / R_S$ and $R_L / R_O$ ratios
on transport opacity and $K_\perp$ for the minijet initial condition.
For $\chi = 0$, both $R_O$ and $R_S$ are determined by the nuclear 
size to be $\approx 3$ fm, therefore $R_O / R_S = 1$.
However, as the opacity increases,
the ratio increases above one for $K_\perp < K_c$, 
while decreases below one for $K_\perp > K_c$,
where the location of the turning point $K_c \approx 1.3-2$ GeV 
depends on the hadronization scheme.

\begin{figure}[hbpt]
\epsfysize=5.5cm
\epsfbox{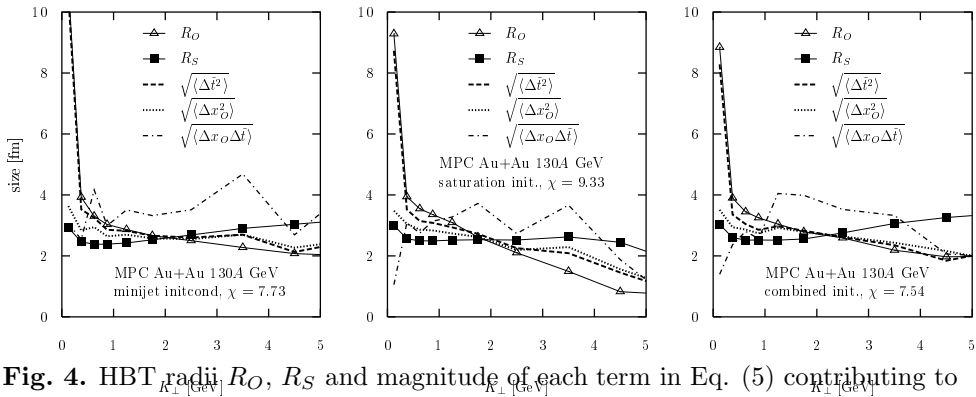}
\vspace*{-1cm}
\caption[]{HBT radii $R_O$, $R_S$ and magnitude of each term
in Eq. (\protect{\ref{HBTradii}}) contributing to $R_O$
as a function of $K_\perp$ for all three classes of initial conditions
with hadronization via parton-hadron duality.}
\label{fig:4}
\end{figure}

The naive $R_O^2 = R_S^2 + \Delta t^2$ expectation does not hold now
because in the LCMS frame the source is not cylindrical (see Fig. \ref{fig:2})
and there are dynamical correlations.
Figure \ref{fig:4} illustrates how the different terms in Eq. (\ref{HBTradii})
contribute to $R_O$ as a function of $K_\perp$. For all three
classes of initial conditions,
the time spread $\langle \Delta \tilde t^2 \rangle $
is comparable to the spatial width of the source $\sim 3$ fm,
which has the effect to increase $R_O$.
However, this effect
is compensated by the large {\em positive} $x_O -\tilde t$ correlation term,
reducing $R_O$ below $R_S$.

The right column in Fig. \ref{fig:3} shows a comparison to the
HBT radii measured at RHIC.
In the transverse opacity range $\chi \sim 0 - 8$ we studied,
the transport theory results are smaller than the observed $R_O$ and $R_L$.
This is in sharp contrast to ideal hydrodynamics,
which overpredicts both radii\cite{Uliradii}.
The monotonic dependence of $R_O$ and $R_L$ on opacity
indicates that a much better agreement with the data is possible
at a larger but {\em finite} opacity $\chi > \sim 10$.

In Fig. \ref{fig:3},
the reason for the anomalous small $R_L$ for zero opacity
is instantaneous freezeout at the formation time $\tau = \tau_0$.
Since $R_L^2 \approx \tau^2 [\Delta(\eta - y)]^2$,
$R_L$ depends on the decoupling time and the strength of the $\eta-y$
correlation.
For our thermally correlated initial condition
$[\Delta(\eta - y)]^2\approx T/m_\perp$ and $\tau_0 = 0.1$ fm$/c$,
which for $\chi = 0$ gives 
$R_L < \approx 0.5$ fm for all $K_\perp$ bins we studied.

However,
as the transport opacity increases,
$R_L$ becomes much larger 
because the decoupling time increases as is evident from Fig. \ref{fig:1}.
The largest increase $\tau/\tau_0 \sim R/\tau_0 \sim 50$
is for low-$p_\perp$ partons, which freeze out latest.
Thus the observed $R_L(K_\perp)$ carries important constraints on the
dynamics of freezeout.
For example, perfect inside-outside correlation, i.e., $\eta = y$
as assumed in classical Yang-Mills approaches
cannot be reconciled with
the RHIC data because it gives $R_L = 0$.

Finally,
the biggest puzzle in Fig. \ref{fig:3}
is why $R_S(K_\perp) \approx const \approx 3$ fm
contrary to the data
and furthermore independent of the transport opacity.
This suggests that $R_S$ is independent of the degree of collective dynamics,
i.e., it is determined solely by the initial nuclear geometry.
The same problem with $R_S$ being too small and insensitive to the dynamical
assumptions
has been observed in hydrodynamical calculations as well\cite{Uliradii}.

\section*{Acknowledgements}
D. M. would like to acknowledge enlightening discussions with Tam\'as Cs\"org\H o and Ulrich Heinz.

This work was supported by the Director, Office of Energy Research,
Division of Nuclear Physics of the Office of High Energy and Nuclear Physics
of the U.S. Department of Energy under contract No. DE-FG-02-93ER-40764.

\vfill\eject

\begin{thebibliography}{99}  
  
\bibitem{starHBT}
C.~Adler {\it et al.}  [STAR Collaboration],
Phys.\ Rev.\ Lett.\  {\bf 87}, 082301 (2001)
[nucl-ex/0107008].

\bibitem{phenixHBT}
K.~Adcox {\it et al.}  [PHENIX Collaboration],
nucl-ex/0201008.

\bibitem{v2}
D.~Molnar and M.~Gyulassy,
Nucl.\ Phys.\ A {\bf 697}, 495 (2002)
[nucl-th/0104073].

\bibitem{ALCOR}
T.~S.~Biro, P.~Levai and J.~Zimanyi,
Phys.\ Lett.\ B {\bf 347}, 6 (1995).

\bibitem{Magas:1999yb}
V.~K.~Magas {\it et al.},
Heavy Ion Phys.\  {\bf 9}, 193 (1999)
[nucl-th/9903045].

\bibitem{Magas:2000jx}
V.~K.~Magas, L.~P.~Csernai and D.~D.~Strottman,
Phys.\ Rev.\ C {\bf 64}, 014901 (2001)
[hep-ph/0010307].

\bibitem{TamasHBT}
T.~Csorgo,
hep-ph/0001233;
A.~Ster, T.~Csorgo and J.~Beier,
Heavy Ion Phys.\  {\bf 10}, 85 (1999)
[hep-ph/9810341].

\bibitem{DirkHBT}
D.~H.~Rischke and M.~Gyulassy,
Nucl.\ Phys.\ A {\bf 608}, 479 (1996)
[nucl-th/9606039].

\bibitem{Uliradii}
U.~W.~Heinz and P.~F.~Kolb,
hep-ph/0111075.

\bibitem{Yang}
Y.~Pang, RHIC 96 Summer Study, CU-TP-815 preprint (unpublished);
Generic Cascade Program (GCP) documentation available at WWW site
http://www-cunuke.phys.columbia.edu/rhic/gcp.

\bibitem{ZPC}
B.~Zhang,
Comput.\ Phys.\ Commun.\  {\bf 109}, 193 (1998)
[nucl-th/9709009].

\bibitem{nonequil}
D.~Molnar and M.~Gyulassy,
Phys.\ Rev.\ C {\bf 62}, 054907 (2000)
[nucl-th/0005051].

\bibitem{finrange}
S.~Cheng {\it et al.},
Phys.\ Rev.\ C {\bf 65}, 024901 (2002)
[nucl-th/0107001].

\bibitem{starv2}
R.~J.~Snellings  [STAR Collaboration],
Nucl.\ Phys.\ A {\bf 698}, 193 (2002)
[nucl-ex/0104006].

\bibitem{hydro+urqmd}
S.~Soff, S.~A.~Bass and A.~Dumitru,
Phys.\ Rev.\ Lett.\  {\bf 86}, 3981 (2001)
[nucl-th/0012085].
 
\bibitem{MPCsource} D.~Moln\'ar, MPC~1.5.9.
This parton cascade code used in the present study
can be downloaded from  WWW at
http://www-cunuke.phys.columbia.edu/people/molnard

\bibitem{MiklosHBT}
M.~Gyulassy, S.~K.~Kauffmann and L.~W.~Wilson,
Phys.\ Rev.\ C {\bf 20}, 2267 (1979).

\bibitem{UliHBT}
U.~A.~Wiedemann and U.~W.~Heinz,
Phys.\ Rept.\  {\bf 319}, 145 (1999)
[nucl-th/9901094].

\bibitem{HIJING}
M.~Gyulassy and X.~Wang,
Comput.\ Phys.\ Commun.\  {\bf 83}, (1994) 307
[nucl-th/9502021].

\SKIP{
\bibitem{BKK95}
J.~Binnewies, B.~A.~Kniehl and G.~Kramer,
Z.\ Phys.\ C {\bf 65}, 471 (1995).
}

\end{thebibliography}
\end{document}